\documentclass[journal]{IEEEtran}

\usepackage[english]{babel}															
\usepackage{graphics} 
\usepackage{arydshln} 
\usepackage{mathtools} 
\usepackage{times} 
\usepackage{amsmath} 
\usepackage{amssymb}  
\usepackage{amsfonts}%
\usepackage{euscript}
\usepackage{mathrsfs}
\usepackage[inline]{enumitem}
\usepackage{xcolor}
\usepackage{subcaption}
\usepackage{cite}
\usepackage{url}
\usepackage{xfrac}
\usepackage[normalem]{ulem}
\usepackage[
  bookmarks=false,
  pdfpagelabels=false,
  hyperfootnotes=false,
  hyperindex=false,
  pageanchor=false,
  implicit=false,
  colorlinks=false,
]{hyperref}
\usepackage{tikz}

\usetikzlibrary{shadows.blur,positioning,backgrounds,fit} 
\usepackage{pgfplots}
\pgfplotsset{compat=newest} 
\pgfplotsset{plot coordinates/math parser=false}
\usepgfplotslibrary{fillbetween}

\newcommand{\NX}{{n_\mathrm{x}}}

\newcommand{\NZ}{{n_\mathrm{z}}}
\newcommand{\NW}{{n_\mathrm{w}}}
\newcommand{\NP}{{n_\mathrm{p}}}

\newcommand{\sP}{\mathbb{P}}
\newcommand{\sX}{\mathbb{X}}

\newcommand{\sZ}{\mathbb{Z}}
\newcommand{\sI}{\mathbb{I}}
\newcommand{\sV}{\mathbb{V}}
\newcommand{\sS}{\mathbb{S}}

\newcommand{\rank}{\mathrm{rank}}

\newcommand{\Afnc}{\mathcal{A}}
\newcommand{\Bfnc}{\mathcal{B}}
\newcommand{\Cfnc}{\mathcal{C}}
\newcommand{\Dfnc}{\mathcal{D}}

\newcommand{\Kfnc}{\mathcal{K}}

\newcommand{\Lfnc}{\mathcal{L}}
\newcommand{\Rfnc}{\mathcal{R}}
\newcommand{\Qfnc}{\mathcal{Q}}

\newcommand{\cnvxhullP}{\mathcal{P}}
\newcommand{\cnvxhullV}{\mathcal{V}}

\newcommand{\conv}{\mathrm{co}}

\newcommand{\pmean}{\mathtt{p}_\mathrm{c}}

\makeatletter
\newcommand{\pushright}[1]{\ifmeasuring@#1\else\omit\hfill$\displaystyle#1$\fi\ignorespaces}
\makeatother

\newtheorem{thm}{Theorem}
\newtheorem{lem}{Lemma}

\newtheorem{defn}{Definition}

\newtheorem{rem}{Remark}

\renewcommand{\IEEEQED}{\IEEEQEDopen}

\begin{document}
\title{Affine Parameter-Dependent Lyapunov Functions \\ for LPV Systems with Affine Dependence}
%
%
%

\author{Pepijn~B.~Cox,~\IEEEmembership{Member,~IEEE,}
        Siep~Weiland,~\IEEEmembership{Member,~IEEE,}
        and~Roland~T\'{o}th,~\IEEEmembership{Senior Member,~IEEE}
\thanks{P.B. Cox, S. Weiland, and R. T\'{o}th are with the Control Systems Group, Department of Electrical Engineering, Eindhoven University of Technology, P.O. Box 513, 5600 MB Eindhoven, The Netherlands, e-mail: \{p.b.cox, s.weiland, r.toth\}@tue.nl.}
\thanks{This paper has received funding from the European Research Council (ERC) under the European Union's Horizon 2020 research and innovation programme (grant agreement No 714663).}}

\ifCLASSOPTIONpeerreview
\markboth{}%
{... \MakeLowercase{\textit{et al.}}: Affine Parameter-Dependent Lyapunov Functions for LPV Systems with Affine Dependence}
\else
\markboth{}%
{Cox \MakeLowercase{\textit{et al.}}: Affine Parameter-Dependent Lyapunov Functions for LPV Systems with Affine Dependence}
\fi


\maketitle

\begin{abstract}
This paper deals with the certification problem for robust quadratic stability, robust state convergence, and robust quadratic performance of linear systems that exhibit bounded rates of variation in their parameters. We consider both continuous-time (CT) and discrete-time (DT) parameter-varying systems. In this paper, we provide a uniform method for this certification problem in both cases and we show that, contrary to what was claimed previously, the DT case requires a significantly different treatment compared to the existing CT results. In the established uniform approach, quadratic Lyapunov functions, that are affine in the parameter, are used to certify robust stability, robust convergence rates, and robust performance in terms of linear matrix inequality feasibility tests. To exemplify the procedure, we solve the certification problem for $\mathscr{L}_2$-gain performance both in the CT and the DT cases. A numerical example is given to show that the proposed approach is less conservative than a method with slack variables.
\end{abstract}


\IEEEpeerreviewmaketitle


\section{Introduction}

\IEEEPARstart{T}{he} certification of stability and performance of a linear system against uncertain and/or time-varying parameters is of paramount interest~\cite{Shamma1988,Mohammadpour2012,Briat2015}. Non-linear and/or non-stationary effects can be represented by varying dynamic behavior of a linear system description expressed as parameter variations. This makes the so-called \textit{linear parameter-varying} (LPV) system representations widely applicable to model physical or chemical processes. However, the certification of robust quadratic stability, robust state convergence, and robust quadratic performance for LPV representations becomes more involved compared to the \emph{linear time-invariant} (LTI) case. There has been extensive research in this area, where the parameter variations are assumed to be either time-invariant in a constraint set or time-varying signals with possible constraints on their values, rates of variations, bandwidths, or spectral contents. For time-invariant parametric dependence and arbitrary fast time-variations, a rich literature exists, e.g., see~\cite{Zhou1996,Ebihara2015,Briat2015}.

In a wide range of applications, systems exhibit bounded parameter variations and bounded rates of variation. For these systems, assumptions on arbitrary fast parameter rates are unrealistic and conservative. Formulation of stability conditions taking the bounded rate of variation into account can be accomplished, e.g., via \textit{integral-quadratic-constraints} (IQCs)~\cite{Megretski1997,Helmersson1999,Powers2001,Molchanov2002,Koroglu2007} based on an equivalent \textit{linear fractional representation} (LFR) of the LPV system, Lyapunov theory based on state-space representations or LFR forms~\cite{Gahinet1996,Chilali1999,Mahmoud2002,Oliveira1999b,Blanchini1999,Daafouz2001,Souza2006,Amato2005}, the quadratic separator theorem for LFRs~\cite{Iwasaki2001,Peaucelle2007,Wei2008}, or Finsler's Lemma~\cite{Oliveira2001}. However, these stability and performance formulations result in an infinite set of \emph{linear matrix inequalities} (LMIs). To obtain tractable methodologies, a number of relaxation techniques can be applied. These include
\begin{enumerate*}[label=\arabic*)]
	\item vertex separation or convex-hull relaxation, including $\mu$ synthesis, $D/G$ scaling, and full-block multipliers,~\cite{Megretski1997,Helmersson1999,Molchanov2002,Koroglu2007,Iwasaki2001,Peaucelle2007,Wei2008};
	\item partial-convexity arguments~\cite{Gahinet1996,Chilali1999,Mahmoud2002};
	\item sum-of-squares relaxation~\cite{Scherer2006,Wei2008};
	\item P\'{o}lya  relaxation~\cite{Oliveira1999b,Blanchini1999,Daafouz2001,Powers2001};
	\item slack variable methods (applied in discrete-time)~\cite{Oliveira1999b,Daafouz2001,Souza2006}; and
	\item partitioning of the uncertainty space~\cite{Shamma1988,Amato2005}.
\end{enumerate*}
The aforementioned overview is far from complete due to the vast amount of literature on this topic. A summary of various relaxation techniques in the \textit{continuous-time} (CT) case is found in~\cite{Briat2015,Veenman2016}.

Many approaches presented in the literature for quadratic stability, state convergence, and quadratic performance certification problems are either conservative or have a high computational demand due to the large number of free parameters or the large number of LMIs. Furthermore, the CT and the \textit{discrete-time} (DT) cases are fundamentally different for these problems. Indeed, in the DT case, the robust stability analysis requires a significantly different relaxation technique leading to more involved synthesis and verification results. By neglecting this difference, the DT case remained mainly unexplored under the claims that it trivially follows from the CT results. We will demonstrate that this is not the case. The goal of this paper is to provide a unified method for certification of robust quadratic stability, robust state convergence, and robust quadratic performance, in both the CT and DT cases, for linear systems that exhibit parameter variations where these variations have bounded rates. The certification problem is solved using parameter-dependent Lyapunov functions. To avoid large numbers of free parameters, to decrease the number of LMIs, and to avoid conservativeness of the results, we apply the partial-convexity relaxation argument inspired by the results in~\cite{Gahinet1996,Chilali1999,Mahmoud2002}.

More specifically, the contributions of the paper are the following, for the CT case, we extend~\cite{Gahinet1996,Chilali1999} to include a certificate on the exponential rate of decay of the state, independent of the uncertain parameter. For the DT case, we extend the partial-convexity argument to handle the third-order terms of the Lyapunov function without introducing additional free parameters and we provide a less conservative robust stability test, a state convergence guarantee, and a performance test than other existing approaches. To the authors' knowledge, the only contribution using the partial-convexity argument in the DT case is~\cite{Mahmoud2002}. However,~\cite[Thm. 1]{Mahmoud2002} avoids the cubic dependencies on the parameters by including additional free parameters and by increasing the number of LMIs. Additionally, we draw a parallel with LMI regions to show why the DT case induces a more involved set of LMIs. In addition, an illustrative example is provided comparing the proposed partial-convexity results with a slack variable based method in the DT case.

The paper is organized as follows. In Section~\ref{sec:LPVsystem}, the concept of LPV systems, state-space representations, and the problem setting are introduced. The analysis on \textit{affine quadratic stability} (AQS) is given in Section~\ref{sec:affineQuadStab}. Section~\ref{sec:suffAQS} reveals how the partial-convexity relaxation of the AQS certification problem results in a finite set of LMIs for the CT and DT cases. In Section~\ref{sec:affineQuaPerf}, an LMI based formulation for the \textit{affine quadratic performance} (AQP) analysis is derived for both cases. In Section~\ref{sec:Example}, an illustrative example is given for the DT case. Concluding remarks are given in Section~\ref{sec:conclusion}.


\section{Linear parameter-varying systems} \label{sec:LPVsystem}

\subsection{LPV state-space representation}

Consider a system defined by the following LPV state-space representation:
\begin{subequations} \label{eq:lpvSys}
\begin{alignat}{3}
  \xi x(t) &= \Afnc(p(t))&x(t)&+\Bfnc(p(t))&&w(t), \label{eq:lpvSysState} \\
  z(t) &= \Cfnc(p(t))&x(t)&+\Dfnc(p(t))&&w(t), \label{eq:lpvSysOut}
\end{alignat}
\end{subequations}
where $x:\mathbb{T}\mapsto\mathbb{R}^\NX$ is the state variable, $p:\mathbb{T}\mapsto\sP\subset\mathbb{R}^\NP$ is the time-varying parameter (i.e., the \textit{scheduling signal}), $w:\mathbb{T}\mapsto\mathbb{R}^\NW$ is the general disturbance channel, $z:\mathbb{T}\mapsto\mathbb{R}^\NZ$ denotes the performance channel, and $t\in\mathbb{T}$ defines time. For the CT case, $\mathbb{T}=\mathbb{R}$ and $\xi=\frac{\mathrm{d}}{\mathrm{d}t}$, i.e., $\xi x(t)=\frac{\mathrm{d}}{\mathrm{d}t}x(t)$ denotes the time derivative of the state; and for the DT case, $\mathbb{T}=\mathbb{Z}$ and $\xi= q$ is the time-shift operator, i.e., $\xi x(t)=x(t+1)$. The parameter-varying matrix functions $\Afnc(\centerdot),\ldots,\Dfnc(\centerdot)$ in~\eqref{eq:lpvSys} are considered to be affine functions of $p$:
\begin{equation}\label{eq:sysMatrices}
\begin{aligned} 
 \Afnc(p(t)\!)&\!=\!A_0\!+\hspace{-1mm}\sum_{i=1}^{\NP} A_ip_i(t),\hspace{-1mm}&\Bfnc(p(t)\!)&\!=\!B_0\!+\hspace{-1mm}\sum_{i=1}^{\NP} B_ip_i(t) , \\
 \Cfnc(p(t)\!)&\!=\!C_0\!+\hspace{-1mm}\sum_{i=1}^{\NP} C_ip_i(t),\hspace{-1mm}&\Dfnc(p(t)\!)&\!=\!D_0\!+\hspace{-1mm}\sum_{i=1}^{\NP} D_ip_i(t),
\end{aligned}
\end{equation}
with known matrices $A_i\in\mathbb{R}^{\NX\times\NX}$, $B_i\in\mathbb{R}^{\NX\times\NW}$, $C_i\in\mathbb{R}^{\NZ\times\NX}$, $D_i\in\mathbb{R}^{\NZ\times\NW}$ for $i=0,\ldots,\NP$ and $p_i$ is the $i$-th element of the scheduling variable. Due to linearity from $x$ to $z$, asymptotic stability \footnote{An LPV system, represented in terms of~\eqref{eq:lpvSys}, is called asymptotically stable, if, for all trajectories of $(w(t),p(t),z(t))$ satisfying~\eqref{eq:lpvSys} with $w(t)=0$ for $t\geq0$ and $p(t)\in\sP$, it holds that $\lim_{t\rightarrow\infty}\vert z(t)\vert = 0$.} of~\eqref{eq:lpvSys} is dictated by stability of the fixed point at the origin of the autonomous part of~\eqref{eq:lpvSys}, i.e.,
\begin{equation} \label{eq:lpvSysAut}
\xi x(t) = \underbrace{\Afnc(p(t))x(t)}_{f(x(t),p(t))}, \hspace{1.5cm} x(0)=x_0,
\end{equation}
where $x_0\in\mathbb{R}^\NX$ denotes the initial state. In this paper, we assume that the scheduling variable and its rate are bounded:
\begin{subequations}
\begin{enumerate}[label={\bf A\arabic*},ref=A\arabic*]
	\item\label{ass:boundedP} The scheduling signal $p(t)$ and its rate of variation $\delta p(t)$, defined by $\delta p(t)=\frac{\mathrm{d}}{\mathrm{d}t} p(t)$ in the CT case and $\delta p(t)= p(t+1)-p(t)$ in the DT case, range for all $t\in\mathbb{T}$ in the bounded set $\sP\times\sV$, i.e., $(p(t),\delta p(t))\in\sP\times\sV$ with $\sV\subset\mathbb{R}^\NP$. Here $\sP=\conv(\cnvxhullP)$ and $\sV=\conv(\cnvxhullV)$ are defined as the convex hulls of the hyper-rectangles
	\begin{alignat}{3}
		\cnvxhullP &= \bigl\{ \bigl[\begin{array}{ccc} u_1&\!\!\!\cdots\!\!\!&u_\NP\end{array}\bigr]^{\!\top}: ~~& u_i\in\{ \underline p_i,  \overline p_i \} \bigr\}, \label{eq:hyper_rectangle_patameter} \\
		\cnvxhullV &= \bigl\{ \bigl[\begin{array}{ccc} v_1&\!\!\!\cdots\!\!\!&v_\NP\end{array}\bigr]^{\!\top}:   &v_i\in\{ \underline\nu_i,  \overline\nu_i \} \bigr\}, \label{eq:hyper_rectangle_patameter_rate}
	\end{alignat}
	with $2^\NP$ vertices each.
\end{enumerate} 
\end{subequations}

\subsection{Notation}
Let $\sI_s^v$ denote the set of integers $\{ s,s+1,\cdots,v\}$ and $\sS^n$ the set of all $n\times n$ real symmetric matrices. The superscript $n$ is omitted if the dimension is not relevant for the context. In addition, the inequalities $A\succeq B$ and $A\succ B$ with $A\in\sS^n$ and $B\in\sS^n$ mean that $A-B$ is positive semi-definite and positive definite, respectively. Let $\pmean$ denote the center of $\sP$:
\begin{equation} \label{eq:schedulingMean}
\pmean = \left[ \begin{array}{ccc} \frac{\underline p_1+\overline p_1}{2} & \ldots & \frac{\underline p_\NP+\overline p_\NP}{2} \end{array} \right]^\top.
\end{equation}
In addition, we will denote multiple summations $\sum_{i=1}^\NP\!\cdots\!\sum_{k=1}^\NP$ by $\sum_{i,\ldots,k=1}^\NP$. $\mathbb{X}^\sZ$ is the standard notation for the collection of all maps from $\sZ$ to $\mathbb{X}$. $\mathbb{R}_{\geq0}$ denotes the set of nonnegative real numbers and $\mathbb{Z}_{\geq0}$ is the set of nonnegative integers. $\mathbb{T}_0=\mathbb{R}_{\geq0}$ in the CT case and $\mathbb{T}_0=\mathbb{Z}_{\geq0}$ in the DT case. With $\sigma_i\{A\}$, we denote the singular values of a real matrix $A$ where the largest and smallest singular values are indicated as $\sigma_\mathrm{max}\{A\}$ and $\sigma_\mathrm{min}\{A\}$, respectively. The eigenvalues of $A$ are denoted as $\lambda_i\{A\}$. The spectral norm of $A$ is defined as $\Vert A \Vert_2=\sigma_\mathrm{max}\{A\}$. Furthermore,  with $\|\!\centerdot\!\|_q$ we denote the vector $q$-norm. The $L_q$-norm of a continuous signal $w$ is $\| w \|_{q}=\left(\int_0^\infty\| w(t) \|^q_q dt\right)^{\sfrac{1}{q}}$ and for a discrete signal $w$, the $\ell_q$-norm is $\| w \|_{q}=\left(\sum_{t=0}^\infty\| w(t) \|^q_q\right)^{\sfrac{1}{q}}$.


\section{Affine quadratic stability} \label{sec:affineQuadStab}

In this paper, asymptotic stability of~\eqref{eq:lpvSysAut}
is verified under~\ref{ass:boundedP} by using a parameter-dependent Lyapunov function.
\begin{defn}[Parameter-dependent Lyapunov function] \label{dfn:lyapunov function}
The function $V:\mathbb{R}^\NX\times\sP\rightarrow\mathbb{R}$ is a parameter-dependent Lyapunov function for~\eqref{eq:lpvSysAut} if:
\begin{enumerate}[label=(\roman*)]
	\item $V(x,p)>0$ for all $x\in\mathbb{R}^\NX$ with $x\neq0$ and $\forall p\in\sP$;
	\item $V(x,p)=0$ for $x=0$ and $\forall p\in\sP$; and
	\item $\Delta V(x,p,r)<0$ for $\forall (x,p,r)\in\mathbb{R}^\NX\times\sP\times\sV$ with $x\neq0$. In the CT case,
	\begin{subequations}
	\begin{equation} \label{defn-eq:LyapunovDerCT}
	\Delta V(x,p,r) \coloneqq \nabla_{\!\!x\!} V(x,p) \cdot\! f(x,p) + \nabla_{\!\!p\!} V(x,p) \cdot r,
	\end{equation}
	where $\nabla_{\!\!y\!}$ indicates the gradient of a function w.r.t. $y$ and $\cdot$ is the inner product. For the DT case,
	\begin{equation} \label{defn-eq:LyapunovDifDT}
	\Delta V(x,p,r) \coloneqq V(f(x,p),p+r) -  V(x,p).
	\end{equation}
	\end{subequations}
\end{enumerate}
\end{defn}

\begin{thm}[Asymptotic stability] \label{thm:AsymptoticStab}
If a parameter-dependent Lyapunov function for~\eqref{eq:lpvSysAut} exists, then the origin of~\eqref{eq:lpvSysAut} is an asymptotically stable fixed point.
\end{thm}
\begin{IEEEproof}
See, for example,~\cite[Thm. 4.1]{Khalil2002} for the CT case and~\cite[Thm. B.23]{Rawlings2015} for the DT case.
\end{IEEEproof}

In this paper, we consider quadratic parameter-dependent Lyapunov functions of the form
\begin{equation} \label{eq:quadParmLyap}
V(x,p) = x^\top \Kfnc(p)x,
\end{equation}
where the function $\Kfnc:\sP\mapsto\sS^\NX$ is affine in $p$
\begin{equation} \label{eq:postiveMatFunc}
\Kfnc(p) = K_0+\sum_{i=1}^\NP K_ip_i,
\end{equation}
with unknown matrices $K_0,\ldots,K_\NP\in\sS^\NX$.

\begin{defn}[Affine quadratic stability]
The representation~\eqref{eq:lpvSys} is \emph{affinely quadratically stable} (AQS), if it admits a parameter-dependent Lyapunov function of the form~\eqref{eq:quadParmLyap}-\eqref{eq:postiveMatFunc}.
\end{defn}

Let us provide sufficient conditions for AQS:
\begin{thm}[Sufficiency for AQS] \label{thm:AQS}
The LPV representation~\eqref{eq:lpvSysAut} is affinely quadratically stable, if there exist $\NP+1$ matrices $K_0,\ldots,K_\NP\in\sS^{\NX}$ such that
\begin{equation}
\Kfnc(p)  = K_0+\sum_{i=1}^\NP K_ip_i \succ 0, \label{thm-eq:LyupunovFunc}
\end{equation}
for all $p\in\sP$ and, in addition, for the CT case, it holds that
\begin{subequations} \label{thm-eq:LyupunovIneq}
\begin{equation}
\Lfnc_\mathrm{c}(p,r)\!=\! \Afnc^{\!\top}\!\!(p) \Kfnc(p)\!+\! \Kfnc(p)\Afnc(p) \!+\! \Kfnc(r)\!-\!K_0 \prec 0, \label{thm-eq:LyupunovIneqCont}
\end{equation}
for all $(p,r)\in\sP\times\sV$ or, for the DT case, it holds that
\begin{equation}
\Lfnc_\mathrm{d}(p,r)\! =\! \Afnc^{\!\top}\!\!(p)\!\left( \Kfnc(p)\! +\!\Kfnc(r)\!-\!K_0\right)\Afnc(p)\! -\! \Kfnc(p) \!\prec\! 0, \label{thm-eq:LyupunovIneqDisc}
\end{equation}
\end{subequations}
for all $(p,r)\in\sP\times\sV$.
\end{thm}
\begin{IEEEproof}
For the CT case, e.g., see~\cite[Thm. 2.4.6]{Briat2015}. The DT case can be obtained by similar arguments.
\end{IEEEproof} 

Theorem~\ref{thm:AQS} results in an infinite set of matrix inequalities in the unknowns $K_i\in\sS^\NX$. In this paper, we convexify~\eqref{thm-eq:LyupunovIneq} using a partial-convexity argument to obtain a tractable solution by a finite number of LMIs. Within the IQC framework, it has been shown that relaxation by partial-convexity for systems in the form~\eqref{eq:lpvSys} lead to a less conservative formulation than a convex-hull relaxation~\cite[p. 9]{Veenman2016}. It is difficult to make generic statements on the potential conservativeness of various relaxation methods. However, the relaxation by partial-convexity results in less decision variables when compared to slack variable relaxation methods as discussed in~\cite{Oliveira1999b,Daafouz2001,Souza2006}. In Section~\ref{sec:Example}, we demonstrate that relaxation by partial-convexity is less conservative than slack variable relaxation. Alternatively, the uncertainty space can be partitioned as in~\cite{Wei2008}, where the grid size leads to a trade-off between a conservative solution and an increased computational burden. Therefore, the partial-convexity argument potentially avoids excessive number of free parameters and decreases the number of LMIs.


\section{Sufficiency for AQS with guaranteed convergence} \label{sec:suffAQS}

\subsection{Relaxation with partial-convexity}

The functions $\Lfnc_\mathrm{c}(p,r)$ in~\eqref{thm-eq:LyupunovIneqCont} and $\Lfnc_\mathrm{d}(p,r)$ in~\eqref{thm-eq:LyupunovIneqDisc} are quadratic and cubic in $p$, respectively, while linear in $r$. Hence, a relaxation technique is required for $p$ only to obtain a finite set of LMIs to verify AQS in Theorem~\ref{thm:AQS}. If the functions $\Lfnc_\mathrm{c}(\centerdot)$ and $\Lfnc_\mathrm{d}(\centerdot)$ are negative definite on $(p,r)\!\in\!\sP\times\sV$, then the maximum of these functions must be negative. Therefore, if it can be ensured that the maximum of $\Lfnc_\mathrm{c}(\centerdot)$ and $\Lfnc_\mathrm{d}(\centerdot)$ is at the vertices $\cnvxhullP$, then~\eqref{thm-eq:LyupunovIneqCont} and~\eqref{thm-eq:LyupunovIneqDisc} reduce to a finite set of LMIs that are required to be satisfied on $\cnvxhullP$ only. The following lemma provides the conditions for this concept:
\begin{lem}[Maximum at the vertices] \label{lem:multconv}
Consider the cubic function $\Lfnc:\sP\mapsto\sS$ defined by
\begin{equation*}
\Lfnc(p) \!=\! Q_0 +\!\! \sum_{i=1}^{\NP} Q_i p_i +\!\! \sum_{i,j=1}^{\NP}\! Q_{i,j} p_ip_j +\!\! \sum_{i,j,k=1}^{\NP} \!\!Q_{i,j,k} p_ip_jp_k,
\end{equation*}
with matrices $Q_i,Q_{j,k},Q_{j,k,l}\in\sS$ for $i\in\sI_0^\NP$, $j,k,l\in\sI_1^\NP$. The function $\Lfnc(\centerdot)$ achieves its maximum at $\cnvxhullP$ if
\begin{equation} \label{lem-eq:confNece}
\frac{1}{2}\frac{\partial^2 \Lfnc(u)}{\partial u_i^2} \!=\! Q_{i,i} \!+\!\! \sum_{j=1}^\NP (Q_{j,i,i}\!+\!Q_{i,j,i}\!+\!Q_{i,i,j}) u_j  \succeq 0,
\end{equation}
for all $(i,u)\!\in\!\sI_1^\NP\!\times\!\cnvxhullP$ with $\sP=\conv(\cnvxhullP)$.
\end{lem}
\begin{IEEEproof}
See Appendix~\ref{app-sec-pf-lem:multconv}.
\end{IEEEproof}
In fact, condition~\eqref{lem-eq:confNece} implies that $\Lfnc(\centerdot)$ is a partial-convex function:
\begin{defn}[Partially convex function] \label{dfn:partialConv}
A twice differentiable function $\Lfnc:\sP\rightarrow\sS$ is \emph{partially convex} if $\sP$ is convex and
\begin{equation} \label{dfn-eq:convFunc}
\frac{\partial^2 \Lfnc(p)}{\partial p_i^2} \succeq 0,\hspace{1cm}\mbox{ for all } (i, p)\in\sI_1^{\NP}\!\times\sP.
\end{equation}
\end{defn}

Note that positive semi-definiteness of~\eqref{dfn-eq:convFunc} is along each independent direction $p_i$ of the scheduling space. This is less demanding than convexity with respect to $p$, which would require the Hessian of $\Lfnc(\centerdot)$ to be positive semi-definite.

By applying the partial-convexity relaxation~\eqref{lem-eq:confNece} in Lemma~\ref{lem:multconv}, the maximum of $\Lfnc_\mathrm{c}(\centerdot)$ and $\Lfnc_\mathrm{d}(\centerdot)$ can be found at the vertices and, therefore, the AQS test in~\eqref{thm-eq:LyupunovFunc}-\eqref{thm-eq:LyupunovIneq} can be reduced to a finite set of LMIs at the vertices only, i.e., $\Lfnc_i(u,v)\prec0$ $\forall (u,v)\in\cnvxhullP\times\cnvxhullV$. This is the core idea introduced in~\cite{Gahinet1996} for the CT case. This concept can also be applied in the DT case, as highlighted in Section~\ref{subsec:DT_AQS}.

By taking into account the cubic terms in $\Lfnc(\centerdot)$, we differ from the approaches presented in~\cite{Gahinet1996,Chilali1999,Mahmoud2002}. These cubic terms are essential in handling the DT case.

\subsection{The discrete-time case} \label{subsec:DT_AQS}

\begin{thm}[Sufficiency for AQS in DT] \label{thm:AQSDiscRate}
Given an LPV system defined by~\eqref{eq:lpvSysAut} with dependency structure~\eqref{eq:sysMatrices} where the scheduling variable $p(t)$ satisfies~\ref{ass:boundedP}.

If there exists an $0<\epsilon<1$, such that the eigenvalues of $\Afnc(\pmean)$ satisfy $\vert\lambda_i(\Afnc(\pmean))\vert < \sqrt{1-\epsilon}$ with $\pmean$ as in~\eqref{eq:schedulingMean}, and there exist $\NP+1$ matrices $K_0,\ldots,K_\NP\in\sS^\NX$ parametrizing $\Kfnc(\centerdot)$ in~\eqref{eq:postiveMatFunc} that satisfy
\begin{align}
&\Lfnc_\mathrm{d}(u,v,\epsilon)\!=\! \Afnc^\top(u) \left[\Kfnc(u) + \Kfnc(v) -K_0 \right] \Afnc(u)   \nonumber \\[1mm]
&	   \hspace{1.5cm}- (1-\epsilon) \Kfnc(u) \preceq 0, \hspace{0.6cm} \forall (u,v)\in\cnvxhullP\times\cnvxhullV, \label{thm-eq:StabilityIneqDiscConv} \\[2mm]
&	 0 \preceq \Afnc^{\!\top}\!\!(u)K_iA_i\!+\!A_i^{\!\top}\! K_i\Afnc(u), \hspace{0.8cm} \forall(i,u)\in\sI_1^\NP\!\times\cnvxhullP, \label{thm-eq:ConveityIneqDiscConv}
\end{align}
then the LPV system corresponding to~\eqref{eq:lpvSysAut} is AQS. In particular, $V(x,p)$ in~\eqref{eq:quadParmLyap} is a Lyapunov function with $\Delta V(x,p,\delta p)\leq -\epsilon V(x,p)$ and the state $x$ robustly converges, i.e., $\Vert x(t)\Vert^2_2\leq \frac{b}{a} (1-\epsilon)^t \Vert x(0)\Vert^2_2$ for any trajectory $(p(t),\delta p(t))\in\sP\times\sV$ for $t\geq0$ where 
\begin{equation} \label{eq:abbounds}
a=\inf_{u\in\cnvxhullP}\lambda_\mathrm{min}\bigl\{\Kfnc(u)\bigr\}, \hspace{1cm} b=\sup_{u\in\cnvxhullP}\lambda_\mathrm{max}\bigl\{\Kfnc(u)\bigr\}. 
\end{equation} 
\end{thm}
\begin{IEEEproof}
The proof has three parts:
\begin{enumerate*}[label=\roman*)]
	\item sufficiency of the conditions for asymptotic stability,
	\item showing that $\Kfnc(p)\succ 0$ for all $p\in\sP$ is implied by $\vert\lambda_i(\Afnc(\pmean))\vert < \sqrt{1-\epsilon}$ together with~\eqref{thm-eq:StabilityIneqDiscConv}-\eqref{thm-eq:ConveityIneqDiscConv} and
	\item convergence of $x(t)$.
\end{enumerate*}
Combining i) and ii) for $0<\epsilon<1$ results in a parameter-dependent Lyapunov function $V(\centerdot)$ according to Theorem~\ref{thm:AQS} and AQS is proven.

\emph{Part i.}	
 The function $\Lfnc_\mathrm{d}(p,\delta p,\epsilon)$ in~\eqref{thm-eq:LyupunovIneqDisc} is affine in $\delta p$ and $\epsilon$, hence, it is sufficient to evaluate~\eqref{thm-eq:LyupunovIneqDisc} at the vertices $\cnvxhullV$ for an $0<\epsilon<1$, i.e.,
\begin{equation} \label{thm-prf-eq:partialConv1}
\Lfnc_\mathrm{d}(p,v,\epsilon) \prec 0, \hspace{0.7cm} \forall p\in\mathbb{P} \mbox{, } \forall v\in\cnvxhullV, \mbox{ and } 0<\epsilon<1.
\end{equation}
To prove AQS, fix $v\in\cnvxhullV$, fix $0<\epsilon<1$, and define $\Rfnc(v)=\Kfnc(v)-K_0$, then we can write~\eqref{thm-prf-eq:partialConv1} as
\begin{align}
&\Lfnc_\mathrm{d}(p,v,\epsilon) = - (1-\epsilon)\Kfnc(p) + A_0^\top \Rfnc(v)A_0 \nonumber \\
&	+ \sum_{i=1}^\NP A_i^\top \Rfnc(v)A_i p_i^2 + \sum_{i=1}^\NP\left(A_0^\top \Rfnc(v)A_i + A_i^\top \Rfnc(v) A_0 \right) p_i \nonumber \\
&	+ \sum_{i,j=1,~i\neq j}^\NP \!\!\!\!\! A_i^\top \Rfnc(v)A_j  p_i p_j + A_0^\top K_0 A_0^\top + \sum_{i=1}^\NP A_i^\top K_iA_ip_i^3 \nonumber \\
&	+ \sum_{i=1}^\NP\left( A_i^\top K_0A_0+A_0^\top K_iA_0+A_0^\top K_0A_i \right)p_i \nonumber \\
&	+ \sum_{i,j=1}^\NP\left( A_i^\top K_jA_0+A_0^\top K_iA_j+A_i^\top K_0A_j \right)p_ip_j \nonumber \\
&	+ \sum_{i,j=1,~i\neq j}^\NP \!\!\!\!\!\left( A_i^\top K_jA_i+A_j^\top K_iA_i+A_i^\top K_iA_j \right) p_i^2p_j \nonumber \\
&	+ \sum_{i,j,k=1,~i\neq j\neq k}^\NP \!\!\!\!\! A^\top_iK_jA_kp_ip_jp_k\prec 0. \label{thm-prf:lyapWriteOut}
\end{align}
Based on the partial-convexity argument, Eq.~\eqref{thm-prf:lyapWriteOut} holds if it holds on $\cnvxhullP$ and the following condition is satisfied
\begin{multline*}
\frac{\partial^2 \Lfnc_\mathrm{d}(p,v,\epsilon)}{\partial p_i^2} = 2 A_i^\top \Rfnc(v)A_i + 6 A_i^\top K_iA_ip_i \\ 
	+ 2 \left( A_i^\top K_iA_0+A_0^\top K_iA_i+A_i^\top K_0A_i \right)+  \\
 	+ 2\!\!\!\sum_{j=1,~i\neq j}^\NP\!\!\!\left( A_i^\top K_jA_i+A_j^\top K_iA_i+A_i^\top K_iA_j \right) p_j \succeq 0,
\end{multline*}
which can be rewritten as
\begin{subequations}
\begin{align}
	\hspace{-0.3cm}\underbrace{A_i^{\!\top}\!\left[ \Kfnc(p)\!+\!\Rfnc(v) \right] A_i}_{\succeq0} + \Afnc^{\!\top}\!(p)K_iA_i\!+\!A_i^{\!\top} K_i\Afnc(p) \succeq&  \label{pr-eq:ConveityIneq1} \\
 	\Afnc^{\!\top}\!(p)K_iA_i\!+\!A_i^\top K_i\Afnc(p) &\succeq 0. \label{pr-eq:ConveityIneq2}
\end{align}
\end{subequations}
Note that the partial-convexity relaxation only requires to be positive semi-definite on the interval $\sP$. Also see that, $A_i^\top\left[ \Kfnc(p)+\Rfnc(v) \right] A_i = A_i^\top\left[ \Kfnc(p+v) \right] A_i $, therefore,~\eqref{pr-eq:ConveityIneq1} to~\eqref{pr-eq:ConveityIneq2} is valid. As a consequence, the partial-convexity argument~\eqref{pr-eq:ConveityIneq2} becomes independent of $v\in\cnvxhullV$. If~\eqref{pr-eq:ConveityIneq2} holds for all $(i,u)\in\sI_1^\NP\!\times\cnvxhullP$ then~\eqref{thm-eq:StabilityIneqDiscConv} is convex on the domain $(p,r)\in\sP\times\sV$ and, by using Lemma~\ref{lem:multconv},~\eqref{thm-eq:StabilityIneqDiscConv} should only be tested on the vertices of the hyper rectangle.

\emph{Part ii.} This part will be proven by contradiction. Since $\delta p\!=\!0$ is admissible with $0\!<\!\epsilon\!<\!1$, we write~\eqref{thm-eq:LyupunovIneqDisc} as
\begin{equation}
 \Afnc^\top(p) \Kfnc(p) \Afnc(p) - \Kfnc(p) \prec 0. \label{eq-prf:minConstraintDisc1}
\end{equation}
In addition, $\sP$ is compact due to \ref{ass:boundedP} and, therefore, $\Kfnc(p)$ is a compact function on $\sP$. Let us assume that $\exists p_0\!\in\!\sP$ such that $\Kfnc(p_0)$ is singular and take $x_0\!\in\!\ker(\Kfnc(p_0))$ with $x_0\neq0$, then
\begin{multline} \label{eq-prf:minConstContraDisc1}
x^\top_0\left[\Afnc^\top(p_0) \Kfnc(p_0) \Afnc(p_0) - \Kfnc(p_0)\right]x_0 = \\ x^\top_0\Afnc^\top(p_0) \Kfnc(p_0) \Afnc(p_0)x_0 \geq 0.
\end{multline}
However,~\eqref{eq-prf:minConstContraDisc1} contradicts~\eqref{eq-prf:minConstraintDisc1}, hence,~\eqref{eq-prf:minConstraintDisc1} ensures that $\Kfnc(p)$ cannot be singular for all admissible $p\in\sP$. As $\vert\lambda_i(\Afnc(\pmean))\vert < \sqrt{1-\epsilon}$, $\Kfnc(\pmean)\succ0$ by the $\mathcal{D}$-stability result~\cite{Chilali1999}. Then, as $\Kfnc(p)$ is compact on $\sP$, by continuity of the eigenvalues of $\Kfnc(p)$,~\eqref{eq-prf:minConstraintDisc1} assures that $aI\preceq \Kfnc(p)\preceq bI$ with $0\!<\!a\!\leq\! b\!<\!\infty$ as defined in~\eqref{eq:abbounds} for all admissible $p\in\sP$ and, therefore,~\eqref{thm-eq:LyupunovFunc} is satisfied.

\emph{Part iii.} We can find an $0\!<\!\epsilon\!<\!1$ such that
\begin{multline}
\Delta V(x,p,\delta p) + \epsilon V(x,p) =  \\
x^\top\left[ \Afnc^\top(p) \Kfnc(p+\delta p)\Afnc(p) - (1-\epsilon) \Kfnc(p) \right] x = \\
x^\top \Lfnc_\mathrm{d}(p,\delta p,\epsilon) x \leq0, \label{eq-pf:expnConvNeasDisc}
\end{multline}
for all $(p,\delta p)\in\sP\times\sV$ and $x(0)\neq 0$. Eq.~\eqref{eq-pf:expnConvNeasDisc} provides that $V(\centerdot)$ is decaying along the solutions of~\eqref{eq:lpvSysAut} according to $V(x(t),p(t)) \leq (1-\epsilon)^tV(x(0),p(0))$ for $t>0$. In addition, $a\Vert x\Vert^2_2\leq V(x,p) \leq b \Vert x\Vert^2_2$ for all $p\in\sP$ and $x\in\sX$ since $\Kfnc(p)$ is positive definite and bounded, hence, we can find that $\Vert x(t)\Vert^2_2\leq \frac{b}{a}(1-\epsilon)^t \Vert x(0)\Vert^2_2$ for $t\geq0$.
\end{IEEEproof}

\begin{rem}
It is an essential part of the proof of Theorem~\ref{thm:AQSDiscRate} (Part i) that the function $\Lfnc_\mathrm{d}(\centerdot)$ in~\eqref{thm-prf:lyapWriteOut} includes third-order terms in $p$. In~\cite[Thm. 1]{Mahmoud2002}, these terms are reduced to second-order by introducing additional variables and coupling constraints. Therefore, the proof and the LMIs in Theorem~\ref{thm:AQSDiscRate} are fundamentally different from~\cite[Thm. 1]{Mahmoud2002}. Moreover, the number of decision variables in~\cite[Thm. 1]{Mahmoud2002} is twice the number of decision variables in Theorem~\ref{thm:AQSDiscRate}.
\end{rem}

To decrease the number of LMI conditions in Theorem~\ref{thm:AQSDiscRate}, a more conservative test for AQS can be derived:
\begin{lem}[Simple AQS in DT] \label{lem:AQS_DT_Simple}
Given an LPV system defined by~\eqref{eq:lpvSysAut} with dependency structure~\eqref{eq:sysMatrices} where the scheduling variable $p(t)$ satisfies~\ref{ass:boundedP}.

If there exist an $0<\epsilon<1$ and $\NP+1$ matrices $K_0,\ldots,K_\NP\in\sS^\NX$ parametrizing $\Kfnc(\centerdot)$ in~\eqref{eq:postiveMatFunc} that satisfy
\begin{align}
\Lfnc_1(u,l,\epsilon) &= \Afnc^\top(u) \Kfnc(l) \Afnc(u) - (1-\epsilon)\Kfnc(u) \prec0, \nonumber \\
		& \hspace{3.5cm}\forall (u,l)\in\cnvxhullP\times\cnvxhullP, \label{thm-eq:StabilityIneq1} \\[1mm]
\Kfnc(u) &\succ 0, \hspace{3.27cm} \forall u\in\cnvxhullP, \label{thm-eq:ConveityIneq1}
\end{align}
then the LPV system corresponding to~\eqref{eq:lpvSysAut} is AQS.  In particular, $V(x,p)$ in~\eqref{eq:quadParmLyap} is a Lyapunov function with $\Delta V(x,p,\delta p)\leq -\epsilon V(x,p)$ and the state $x$ robustly converges as $\Vert x(t)\Vert^2_2\leq \frac{b}{a} (1-\epsilon)^t \Vert x(0)\Vert^2_2$ for any trajectory $(p(t),\delta p(t))\in\sP\times\sV$ for $t\geq0$ with $a$ and $b$ given by~\eqref{eq:abbounds}.
\end{lem}
\begin{IEEEproof}
Note that~\eqref{thm-eq:StabilityIneq1} is equivalent to~\eqref{thm-eq:LyupunovIneqDisc}, however, in~\eqref{thm-eq:StabilityIneq1}, $qp\in\sP$ and $p\in\sP$ are treated as independent variables, which implies that $\underline\nu_i= \underline p_i- \overline p_i$ and $\overline\nu_i= \overline p_i-\underline p_i$ for $i\in\sI_1^\NP$. Note that the function $\Lfnc_1(p,qp,\epsilon)$ for $p,qp\in\sP$, $0<\epsilon<1$ is affine in $qp$ and $\epsilon$, hence~\eqref{thm-eq:StabilityIneq1} holds if and only if
\begin{equation}
\Lfnc_1(p,l,\epsilon) \prec 0, \hspace{0.7cm} \forall p\in\sP \mbox{, } \forall l\in\cnvxhullP,\mbox{ and } 0<\epsilon<1.
\end{equation}
Fix $l$ and $\epsilon$, the partial-convexity on the domain $p\in\sP$ gives
\begin{equation} \label{thm-eq:MultConv}
\frac{\partial^2 \Lfnc_1(p,l,\epsilon)}{\partial p_i^2} = A^\top_i \Kfnc(l) A_i \succeq 0, \hspace{0.5cm} \forall (i,l)\in\sI_1^\NP\times\cnvxhullP.
\end{equation}
As~\eqref{thm-eq:ConveityIneq1} enforces $\Kfnc(l)\succ0$,~\eqref{thm-eq:MultConv} is satisfied. Hence, satisfying~\eqref{thm-eq:StabilityIneq1} and~\eqref{thm-eq:ConveityIneq1} results in AQS. Robust state convergence can be proven similarly to Part iii of the proof of Theorem~\ref{thm:AQSDiscRate} and is therefore omitted.
\end{IEEEproof}

Note that Lemma \ref{lem:AQS_DT_Simple} does not depend on the rate of variation $\delta p$, but we implicitly assume that $\delta p$ cannot exceed the difference between the extremes of $\sP$, i.e., $\underline\nu_i= \underline p_i- \overline p_i$ and $\overline\nu_i= \overline p_i-\underline p_i$ for $i\in\sI_1^\NP$. This approach can only be exploited in the DT setting. The number of LMIs is decreased, however, the feasibility test is more conservative compared to Theorem~\ref{thm:AQSDiscRate} as the range $[\underline\nu_i,~\overline\nu_i]$ is enlarged.

\subsection{Generalization}

\begin{thm}[AQS in CT and DT with guaranteed decay] \label{thm:AQSGen}
Given an LPV system defined by~\eqref{eq:lpvSysAut} with dependency structure~\eqref{eq:sysMatrices} where the scheduling variable $p(t)$ satisfies~\ref{ass:boundedP}.

Define the following set of inequalities
\begin{align}
\Lfnc_k(u,v,\alpha)&\!=\!\!\left[\!\!\begin{array}{c} I \\ \Afnc(u) \end{array}\!\! \right]^{\!\!\top}\!\!\!\!\! \bigl(M_k(\alpha)\! \otimes\!  \Kfnc(u) \bigr) \!\!\! \left[\!\!\begin{array}{c} I \\ \Afnc(u)  \end{array} \!\!\right] \!\!+\! \Qfnc_k(u,v) \!\preceq\! 0, \nonumber \\
& \hspace{3.1cm} \forall (u,v)\in\cnvxhullP\times\cnvxhullV,  \label{thm-eq:StabilityIneq} \\[1.5mm]
\mathcal{N}_k(i,u,\alpha) &\!=\! m^{[k]}_{22}\!\left(\!A^{\!\top}\!(u)K_iA_i\!+\!A^{\!\top}_i\!K_i\Afnc(u)\!\right) \!+\! m^{[k]}_{12}K_iA_i+ \nonumber\\
& \hspace{0.4cm} m^{[k]}_{21}A^\top_iK_i \succeq 0,  \hspace{0.3cm} \forall (i,u)\in\sI_1^\NP\!\times\cnvxhullP, \label{thm-eq:ConveityIneq}
\end{align}
where $k\!\in\!\{\mathrm{c},\mathrm{d}\}$ and  $m^{[k]}_{ij}$ is the $i,j$-th element of $M_k(\alpha)$.

For the CT case with $k=\mathrm{c}$, if there exists an $\alpha>0$ such that $\mathrm{Re}\left(\lambda_i(\Afnc(\pmean))\right)<-\alpha$ with $\pmean$ as in~\eqref{eq:schedulingMean} and there exist $\NP+1$ matrices $K_0,\ldots,K_\NP\in\sS^\NX$ parametrizing $\Kfnc(\centerdot)$ in~\eqref{eq:postiveMatFunc} such that~\eqref{thm-eq:StabilityIneq}-\eqref{thm-eq:ConveityIneq} are satisfied with
\begin{equation}
M_{\mathrm{c}}(\alpha) \coloneqq \left[ \begin{array}{cc} 2\alpha & 1 \\ 1 & 0 \end{array} \right], \hspace{0.5cm} \Qfnc_\mathrm{c}(u,v) \coloneqq \Kfnc(v)-K_0, \label{thm-eq:StabilityVariableCont}
\end{equation}
then the LPV system corresponding to~\eqref{eq:lpvSysAut} is AQS. Moreover, $V(x,p)$ in~\eqref{eq:quadParmLyap} is a Lyapunov function with $\Delta V(x,p,\delta p)\leq -2\alpha V(x,p)$ and the state $x$ robustly converges as $\Vert x(t)\Vert^2_2\leq \frac{b}{a}e^{-\alpha t} \Vert x(0)\Vert^2_2$ for any trajectory $(p(t),\delta p(t))\in\sP\times\sV$ for $t\geq0$ with $a$ and $b$ given by~\eqref{eq:abbounds}.

For the DT case with $k=\mathrm{d}$, if there exists an $0<\alpha<1$ such that $\left|\lambda_i(\Afnc(\pmean))\right|< \alpha$ with $\pmean$ as in~\eqref{eq:schedulingMean} and there exist $\NP+1$ matrices $K_0,\ldots,K_\NP\in\sS^\NX$ parametrizing $\Kfnc(\centerdot)$ in~\eqref{eq:postiveMatFunc} such that~\eqref{thm-eq:StabilityIneq}-\eqref{thm-eq:ConveityIneq} are satisfied with
\begin{equation}
M_{\mathrm{d}}(\alpha) \!\coloneqq\!\! \left[\!\! \begin{array}{cc} -\alpha^2 & 0 \\ 0 & 1 \end{array}\! \right]\!\!, ~~\Qfnc_\mathrm{d}(u,v)\! \coloneqq\! \Afnc^{\!\top}\!\!(u)\!\!\left[\Kfnc(v)\!-\!K_0\right]\!\Afnc(u), \label{thm-eq:StabilityVariableDisc}
\end{equation}
then the LPV system corresponding to~\eqref{eq:lpvSysAut} is AQS. In particular, $V(x,p)$ in~\eqref{eq:quadParmLyap} is a Lyapunov function with $\Delta V(x,p,\delta p)\leq -\alpha^2 V(x,p)$ and the state $x$ robustly converges as $\Vert x(t)\Vert^2_2\leq \frac{b}{a} \alpha^{2t} \Vert x(0)\Vert^2_2$ for any trajectory $(p(t),\delta p(t))\in\sP\times\sV$ for $t\geq0$ with $a$ and $b$ given by~\eqref{eq:abbounds}.
\end{thm}
\begin{IEEEproof}
For the CT case, AQS is proven in~\cite[Thm. 3.2]{Gahinet1996} while robust state convergence is proven similar to \cite[Prop. 5.6]{Scherer2015}. For the DT case, see Theorem~\ref{thm:AQSDiscRate}.
\end{IEEEproof}

\begin{rem}
Theorem~\ref{thm:AQSGen} can also be applied for stability analysis under the assumption of a time-invariant scheduling signal, i.e., $\mathbb{V}=\cnvxhullV=\emptyset$ (robust analysis). In this simplification, all elements in~\eqref{thm-eq:ConveityIneq} with respect to the time-variation become zero, i.e., $\Qfnc_\mathrm{c}(\centerdot)\!=\!\Qfnc_\mathrm{d}(\centerdot)\!=\!0$ in~\eqref{thm-eq:StabilityVariableCont} or~\eqref{thm-eq:StabilityVariableDisc}.
\end{rem}

\begin{rem}
Theorem~\ref{thm:AQSGen} implies that, for the CT case, the real parts of the eigenvalues of $\Afnc(p)$ are negative for fixed values of $p$, i.e., $\mathrm{Re}\left(\lambda_i(\Afnc(p))\right)<-\alpha$ $\forall p\in\mathbb{P}$ and, for the DT case, the eigenvalues are within a disc of radius $\alpha$, i.e., $\vert\lambda_i(\Afnc(p))\vert < \alpha$, $\forall p\in\mathbb{P}$. This connects to well-known results on LMI regions, e.g., see~\cite{Chilali1999}. Contrary to the LTI case, the location of the eigenvalues are a necessary, but not sufficient condition for stability in the parameter-varying case, as the contribution of $\Qfnc_k(\centerdot)$ in~\eqref{thm-eq:StabilityIneq} cannot be neglected.
\end{rem}


\section{Affine Quadratic Performance} \label{sec:affineQuaPerf}

\subsection{Concept of dissipativity and performance}

The results on AQS of Section~\ref{sec:suffAQS} can be extended to quadratic performance measures, including $\mathscr{L}_2$ performance, positivity, and $\mathcal{H}_2$ performance.

The LPV system corresponding to~\eqref{eq:lpvSys} is \textit{strictly dissipative} for a given supply function $s:\mathbb{R}^\NW\times\mathbb{R}^\NZ\rightarrow\mathbb{R}$ if there exist an $\varepsilon>0$ and a storage function $V(\centerdot)$ such that,
\begin{equation} \label{eq:strictDisspative}
\Delta V(x,p,\delta p) \leq s(w,z) - \varepsilon \Vert w\Vert_2^2,
\end{equation}
for all $(w,p,\delta p,x,z)\in\mathbb{R}^\NW\times\sP\times\sV\times\mathbb{R}^\NX\times\mathbb{R}^\NZ$, e.g., see~\cite{Scherer2015}. In other words, the change of internal storage $\Delta V(x,p, \delta p)$ will never exceed the amount of supply $s(\centerdot)$ that flows into the system. The inequality~\eqref{eq:strictDisspative} should be satisfied in a point-wise fashion, implying that~\eqref{eq:strictDisspative} also holds for all admissible trajectories that satisfy~\eqref{eq:lpvSys}. Additionally, we take $s(\centerdot)$ to be the following quadratic supply function
\begin{equation} \label{eq:quadStoreFunc}
s(w,z)=\left[\begin{array}{c} w \\ z \end{array} \right]^\top\!\!\!\! P_\mathrm{s}\left[\begin{array}{c} w \\ z \end{array} \right],\hspace{0.75cm} P_\mathrm{s} = \left[\begin{array}{cc} Q_\mathrm{s} & S_\mathrm{s} \\ S^\top_\mathrm{s} & R_\mathrm{s} \end{array} \right],
\end{equation}
where $P_\mathrm{s}\in\sS^{\NW+\NZ}$ is partitioned as $Q_\mathrm{s}\in\sS^{\NW}$, $S_\mathrm{s}\in\mathbb{R}^{\NW\times\NZ}$, and $R_\mathrm{s}\in\sS^{\NZ}$. By appropriately parametrizing~\eqref{eq:quadStoreFunc}, various performance measures can be represented. We test feasibility of~\eqref{eq:strictDisspative} with an affine parameter-dependent function~\eqref{eq:quadParmLyap}, hence, $V(\centerdot)$ qualifies as a quadratic storage function for~\eqref{eq:lpvSys} and~\eqref{eq:quadStoreFunc}. We say that the system~\eqref{eq:lpvSys} achieves \textit{affine quadratic performance} (AQP) for~\eqref{eq:quadStoreFunc} whenever such a quadratic storage function exists.

\subsection{$\mathscr{L}_2$-Gain performance}

To simplify notation, we use $\mathscr{L}_2$ to indicate $L_2$ in CT and $\ell_2$ in DT.

\begin{defn}[Induced $\mathscr{L}_2$-gain] \label{def:L2perf}
Any finite $\gamma$ that satisfies 
\begin{equation} \label{eq:ell2Gain}
\sup_{\substack{0<\| w\|_{2}<\infty \\ p\in\sP^{\mathbb{T}_0}}} \frac{\| z\|_{2}}{\| w\|_{2}}<\gamma,
\end{equation}
is an $\mathscr{L}_2$-gain upper bound where $z:\mathbb{T}_0\rightarrow\mathbb{R}^\NZ$ is the response (performance variable) of~\eqref{eq:lpvSys} for $t\geq0$ under $x_0=0$, general disturbance $w\in L_2(\mathbb{R}^\NW,\mathbb{T}_0)$ in CT or $w\in \ell_2(\mathbb{R}^\NW,\mathbb{T}_0)$ in DT, and scheduling $p\in\sP^{\mathbb{T}_0}$.
\end{defn}

The $\mathscr{L}_2$-gain performance measure of Definition~\ref{def:L2perf} can be characterized by the supply function~\eqref{eq:quadStoreFunc} by choosing  $Q_\mathrm{s} =\gamma^2 I $, $S_\mathrm{s} = 0$, $R_\mathrm{s}=-I$, and $\varepsilon>0$~\cite[Prop. 3.12]{Scherer2015}.

\begin{lem}[$\mathscr{L}_2$-gain performance] \label{lem:AQP}
Given an LPV system defined by~\eqref{eq:lpvSys} with dependency structure~\eqref{eq:sysMatrices} where the scheduling variable $p(t)$ satisfies~\ref{ass:boundedP}.

If there exist a $\gamma>0$ and $\NP+1$ matrices $K_0,\ldots,K_\NP\in\sS^\NX$ parametrizing $\Kfnc(\centerdot)$ in~\eqref{eq:postiveMatFunc} such that
\begin{align}
\Kfnc(p)   &\succ 0, &\forall p&\!\in\!\sP, \label{dfn-eq:LyupunovFuncHinf} \\
\left[\!\!\! \begin{array}{ccc}
\Lfnc_k(p,r) 					& \Kfnc(p) \Bfnc(p) 	& \!\!\Cfnc^{\!\top}\!\!(p) \\
\Bfnc^{\!\top}\!\!(p) \Kfnc(p)\!\!  	& -\gamma I			& \!\!\Dfnc^{\!\top}\!\!(p) \\
\Cfnc(p) 						& \Dfnc(p)			& \!\!-\gamma I
\end{array}\!\!\! \right] \!&\prec\! 0,  &\forall (p,r)&\!\in\!\sP\!\times\!\sV,\label{def-eq:hinfPer}
\end{align}
where $\Lfnc_k(p,r)$ as in~\eqref{thm-eq:LyupunovIneq}, then the LPV system represented by~\eqref{eq:lpvSys} is AQS and has an $\mathscr{L}_2$-gain performance bound $\gamma$.
\end{lem}

\begin{IEEEproof}
See that the $(1,1)$-block of~\eqref{def-eq:hinfPer} implies AQS (Lemma~\ref{thm:AQSGen}). In addition, for $\varepsilon>0$, Eq.~\eqref{def-eq:hinfPer} ensures that
\begin{equation}
\Delta V(x,p,\delta p) + z^\top z - \gamma^2 w^\top w < 0,
\end{equation}
for all admissible $(w,p,\delta p,r,x,z)$. The choice of supply function implies the $\mathscr{L}_2$-gain performance bound of Definition~\ref{def:L2perf} based on similar arguments as in~\cite[Thm. 5.16]{Scherer2015}.
\end{IEEEproof}

As~\eqref{dfn-eq:LyupunovFuncHinf} and~\eqref{def-eq:hinfPer} impose an infinite number of LMIs, we make use of the partial-convexity argument to find an LMI-based test for AQP.

\begin{thm}[Sufficiency for affine $\mathscr{L}_2$-gain performance]
Given an LPV system defined by~\eqref{eq:lpvSys} with dependency structure~\eqref{eq:sysMatrices} where the scheduling variable $p(t)$ satisfies~\ref{ass:boundedP}.

Define the following set of inequalities and equalities %
\footnote{In terms of condition~\eqref{thm-eq:LinfInqCon-eq}, if any $B_i$ for $i=1,\ldots,\NP$ is full row rank then the corresponding $K_i$ must be zero, hence, it can be removed from~\eqref{thm-eq:LinfInq} and~\eqref{thm-eq:LinfInqCon}. If all $B_i$ are full row rank, then we obtain a parameter-independent Lyapunov function.}
\begin{equation} \label{thm-eq:LinfInq}
\left[ \!\!\!\begin{array}{ccc}
\Lfnc_k(u,v,\alpha) 	& \!\Kfnc(u) \Bfnc(u)\! 		& \!\!\Cfnc^{\!\top}\!\!(u) \\
\Bfnc^{\!\top}\!\!(u) \Kfnc(u)\!\!  	& -\gamma I				&\!\! \Dfnc^\top(u) \\
\Cfnc(u) 				& \Dfnc(u)				& \!\!-\gamma I
\end{array}\!\!\! \right] \!\!\preceq\! 0,~  \forall(u,v)\!\in\!\cnvxhullP\!\times\!\cnvxhullV,
\end{equation} \vskip -5mm \noindent
\begin{subequations} \label{thm-eq:LinfInqCon}
\begin{align}
\mathcal{N}_k(u,i,\alpha) &\succeq 0, &  \forall(i, u)&\in\sI_1^\NP\!\times\!\cnvxhullP, \hspace{-0.74cm} \label{thm-eq:LinfInqCon-inq} \\[0mm]
K_iB_i &= 0, &\forall i&\in\sI_1^\NP, \hspace{-0.74cm} \label{thm-eq:LinfInqCon-eq}
\end{align}
\end{subequations}
where $\Lfnc_k(\centerdot)$ and $\mathcal{N}_k(\centerdot)$ are defined in~\eqref{thm-eq:StabilityIneq} and~\eqref{thm-eq:ConveityIneq}. 

For the CT case with $k=\mathrm{c}$, if there exists an $\alpha>0$ such that $\mathrm{Re}\left(\lambda_i(\Afnc(\pmean))\right)<-\alpha$ with $\pmean$ as in~\eqref{eq:schedulingMean} and there exist $\NP+1$ matrices $K_0,\ldots,K_\NP\in\sS^\NX$ such that~\eqref{thm-eq:LinfInq}-\eqref{thm-eq:LinfInqCon} are satisfied, then the LPV system corresponding to~\eqref{eq:lpvSys} is AQS and has an $L_2$-gain performance bound $\gamma$. Moreover, $V(x,p)$ in~\eqref{eq:quadParmLyap} is a Lyapunov function with $\Delta V(x,p,\delta p)\leq -2\alpha V(x,p)$ and the state $x$ robustly converges as $\Vert x(t)\Vert^2_2\leq \frac{b}{a}e^{-\alpha t} \Vert x(0)\Vert^2_2$ for any trajectory $(p(t),\delta p(t))\in\sP\times\sV$ for $t\geq0$ with $w(t)=0$ and $a$, $b$ given by~\eqref{eq:abbounds}.

For the DT case with $k=\mathrm{d}$, if there exists an $0<\alpha<1$ such that $\left|\lambda_i(\Afnc(\pmean))\right|< \alpha$ with $\pmean$ as in~\eqref{eq:schedulingMean} and there exist $\NP+1$ matrices $K_0,\ldots,K_\NP\in\sS^\NX$ such that~\eqref{thm-eq:StabilityIneq}-\eqref{thm-eq:ConveityIneq} are satisfied, then the LPV system corresponding to~\eqref{eq:lpvSys} is AQS with $\ell_2$-gain performance bound $\gamma$. In particular, $V(x,p)$ in~\eqref{eq:quadParmLyap} is a Lyapunov function with $\Delta V(x,p,\delta p)\leq -\alpha^2 V(x,p)$ and the state $x$ robustly converges as $\Vert x(t)\Vert^2_2\leq \frac{b}{a} \alpha^{2t} \Vert x(0)\Vert^2_2$ for any trajectory $(p(t),\delta p(t))\in\sP\times\sV$ for $t\geq0$ with $w(t)=0$ and $a$, $b$ given by~\eqref{eq:abbounds}.
\end{thm}
\begin{IEEEproof}
Sufficiency for affine $\mathscr{L}_2$-gain performance is obtained with robust state convergence if~\eqref{thm-eq:LinfInq} holds together with the following set of inequalities:
\begin{equation}
\left[ \begin{array}{cc}
\mathcal{N}_k(u,i,\alpha) & K_iB_i \\
B_i^\top K_i & 0		
\end{array} \right] \succeq 0, \mbox{ ~for all } (i, u)\in\sI_1^\NP\!\times\cnvxhullP. \label{thm-prf-eq:LinfInqCon}
\end{equation}
This statement can be proven by straightforward application of Theorem~\ref{thm:AQSGen} and Lemma~\ref{lem:AQP}. Hence, in the remainder of the proof it is shown that~\eqref{thm-prf-eq:LinfInqCon} is equivalent to~\eqref{thm-eq:LinfInqCon}.

For the inequality~\eqref{thm-prf-eq:LinfInqCon} to hold, the blocks on the diagonal need to be positive semi-definite, i.e., $\mathcal{N}_k(u,i,\alpha)\succeq0$. The matrix $\mathcal{N}_k(\centerdot)$ might be rank deficient, i.e., $\rank(\mathcal{N}_k(\centerdot)) = n_\mathcal{N}\leq\NX$. Hence, after an appropriate congruence transform on~\eqref{thm-prf-eq:LinfInqCon}, we obtain the following partition of~\eqref{thm-prf-eq:LinfInqCon}: 
\begin{equation} \label{thm-prf-eq:LinfInqCon-step1}
\left[ \!\begin{array}{cc:c}
\widetilde{\mathcal{N}}_k(u,i,\alpha)\!\!\!\! &\!\! 0\!\! & \!\!M_{i,1} \\
0 & \!\!0\!\! & \!\!M_{i,2} \\ \cdashline{1-3} \rule{0pt}{2.6ex}
M_{i,1}^\top\!\! & \!\!M_{i,2}^\top  & \!\!0		
\end{array} \!\right] \!\succeq \!0 \mbox{ with } \widetilde{\mathcal{N}}_k(u,i,\alpha)\!\succ\! 0,
\end{equation}
where $\widetilde{\mathcal{N}}_k(u,i,\alpha)\in\sS^{n_\mathcal{N}}$ and $[M_{i,1}^\top~ M_{i,2}^\top]^\top$ is the partition of $K_iB_i$ after column and row re-arrangement with $M_{i,1}\in\mathbb{R}^{n_\mathcal{N}\times\NW}$, $M_{i,2}\in\mathbb{R}^{\NX-n_\mathcal{N}\times\NW}$. Note that, in case $\mathcal{N}_k(\centerdot)$ is positive definite, $\widetilde{\mathcal{N}}_k(\centerdot)=\mathcal{N}_k(\centerdot)$, $M_{i,1}=K_iB_i$, and $M_{i,2}=\emptyset$. Then, perform a Schur transform on~\eqref{thm-prf-eq:LinfInqCon-step1} to obtain that~\eqref{thm-prf-eq:LinfInqCon-step1} is equivalent to
\begin{equation} \label{thm-pf-eq:AQPSchurPart}
\left[\!\!\! \begin{array}{cc} 0\!\!\! & M_{i,2} \\  M_{i,2}^\top\!\!\!\!  & - M_{i,1}^\top\widetilde{\mathcal{N}}^{-1}_k\!(u,i,\alpha)M_{i,1} \end{array}\!\!\! \right] \!\!\succeq\! 0,~~ \tilde{\mathcal{N}}_{\!k}(u,i,\alpha)\!\succ\! 0.
\end{equation}
Then~\eqref{thm-pf-eq:AQPSchurPart} implies that the blocks on the diagonal are required to be positive semi-definite, i.e., $-M_{i,1}^\top\widetilde{\mathcal{N}}^{-1}_k(\centerdot)M_{i,1}\succeq 0 $. As $\widetilde{\mathcal{N}}_k(\centerdot)$ is positive definite, the latter condition is met if and only if:
\begin{equation*}
M_{i,1}^\top\widetilde{\mathcal{N}}^{-1}_k(u,i,\alpha)M_{i,1}= 0 \hspace{0.5cm} \Longleftrightarrow \hspace{0.5cm} M_{i,1}=0.
\end{equation*}
Then, it follows from~\eqref{thm-pf-eq:AQPSchurPart} that $M_{i,2}=0$. Next, the matrix $K_iB_i$ can be reconstructed from elementary matrix operations based on $M_{i,1}$ and $M_{i,2}$ and we can conclude that $K_iB_i=0$ (undo partition~\eqref{thm-prf-eq:LinfInqCon-step1}).
\end{IEEEproof}
To obtain the minimum of $\gamma$ satisfying~\eqref{eq:ell2Gain}, we solve $\min_{\gamma\geq0}~ \gamma$ s.t. \eqref{thm-eq:LinfInq}-\eqref{thm-eq:LinfInqCon} hold. This convex optimization problem can be solved efficiently by numerical LMI solvers.


\section{Numerical example} \label{sec:Example}

\begin{figure}
	\input{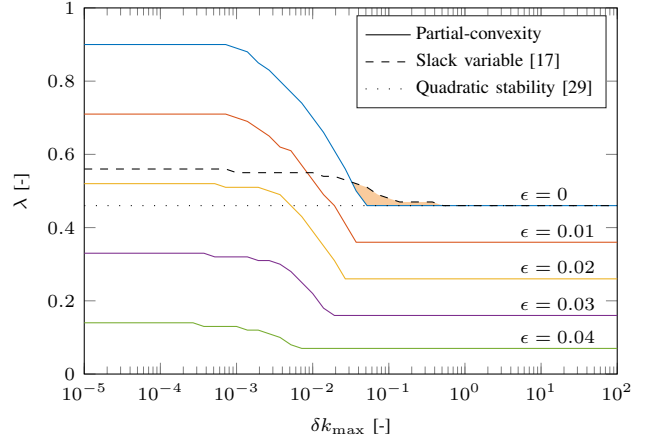}
	\caption{Maximum of $\lambda$ to guarantee AQS w.r.t. $\lambda\sP$ under time-invariant $c$ ($\delta c_\mathrm{max}=0$) and time-varying $k$, where $\epsilon$ indicates the guaranteed rate of convergence of the state, i.e., $\Vert x_t\Vert^2_2\leq \frac{b}{a} (1-\epsilon)^t \Vert x_0\Vert^2_2$ for all $t>0$. The orange area indicates numerical problems with the partial-convexity method using $\epsilon=0$.} \label{fig:exp2_DT}
\end{figure}

\begin{figure}
%
\definecolor{mycolor1}{rgb}{0.00000,0.44700,0.74100}%
\definecolor{mycolor2}{rgb}{0.85000,0.32500,0.09800}%
\definecolor{mycolor3}{rgb}{0.92900,0.69400,0.12500}%
\definecolor{mycolor4}{rgb}{0.49400,0.18400,0.55600}%
\definecolor{mycolor5}{rgb}{0.46600,0.67400,0.18800}%
\definecolor{mycolor6}{rgb}{0.30100,0.74500,0.93300}%
\definecolor{mycolor7}{rgb}{0.63500,0.07800,0.18400}%
\begin{tikzpicture}[font=\scriptsize]

\begin{axis}[%
width=0.8\columnwidth,
height=0.55\columnwidth,
at={(0\columnwidth,0\columnwidth)},
scale only axis,
unbounded coords=jump,
xmode=log,
xmin=1e-05,
xmax=100,
xminorticks=true,
xlabel={$\delta k_\mathrm{max}$ [-]},
ymode=log,
ymin=1,
ymax=300,
yminorticks=true,
ylabel={$\gamma$ [-]},
legend style={legend cell align=left,align=left,draw=white!15!black},
]
\addplot [color=mycolor1,solid,forget plot]
  table[row sep=crcr]{%
1e-05	1.64456063904932\\
1.95734178148766e-05	1.64477170197394\\
3.83118684955728e-05	1.6451842128416\\
7.49894209332456e-05	1.64599296223445\\
0.000146779926762207	1.64757556851752\\
0.000287298483335366	1.65066937608469\\
0.000562341325190349	1.65671717672223\\
0.00110069417125221	1.6685298524308\\
0.00215443469003188	1.69150256554126\\
0.00421696503428582	1.73596046577835\\
0.00825404185268018	1.82079908472308\\
0.0161559809843987	1.96844548763898\\
0.0316227766016838	2.02450651422894\\
0.061896581889126	2.02450630884033\\
0.121152765862859	2.02450686682736\\
0.237137370566166	2.02450613641985\\
0.464158883361278	2.02450610575528\\
0.908517575651686	2.02450591182004\\
1.77827941003892	2.02450573369394\\
3.48070058842841	2.02450566274698\\
6.81292069057961	2.02450637686926\\
13.3352143216332	2.02450587366255\\
26.1015721568254	2.02450497185852\\
51.0896977450692	2.02450592058041\\
100	2.02450598417916\\
};
\node[black,anchor=south west] at (axis cs:4e0,1.85){$\epsilon=0$};
\addlegendimage{line legend,black}
\addlegendentry{Partial-convexity}

\addplot [color=black,dashed]
  table[row sep=crcr]{%
1e-05	1.7267328031792\\
1.95734178148766e-05	1.7268366621332\\
3.83118684955728e-05	1.72702801924328\\
7.49894209332456e-05	1.72709029090371\\
0.000146779926762207	1.727840082481\\
0.000287298483335366	1.72953362531236\\
0.000562341325190349	1.73230511086671\\
0.00110069417125221	1.73803465777932\\
0.00215443469003188	1.74843404331917\\
0.00421696503428582	1.7685673635301\\
0.00825404185268018	1.80506793290015\\
0.0161559809843987	1.86187604373555\\
0.0316227766016838	1.92813879975718\\
0.061896581889126	1.9731466871357\\
0.121152765862859	1.99013728673234\\
0.237137370566166	1.99501345454747\\
0.464158883361278	1.99632348099575\\
0.908517575651686	1.99665437077872\\
1.77827941003892	1.99672722840155\\
3.48070058842841	1.99673497903614\\
6.81292069057961	1.99672999511518\\
13.3352143216332	1.99672494372362\\
26.1015721568254	1.99672171400378\\
51.0896977450692	1.99671990239552\\
100	1.99671893686825\\
};
\addlegendentry{Slack variable~\cite{Souza2006}};

\addplot [color=mycolor2,solid, forget plot]
  table[row sep=crcr]{%
1e-05	2.3009218374147\\
1.95734178148766e-05	2.30147455297165\\
3.83118684955728e-05	2.30218014323456\\
7.49894209332456e-05	2.3039543146174\\
0.000146779926762207	2.30745409325797\\
0.000287298483335366	2.3143426081468\\
0.000562341325190349	2.32782141919927\\
0.00110069417125221	2.35407612157467\\
0.00215443469003188	2.40753585280196\\
0.00421696503428582	2.51262414603874\\
0.00825404185268018	2.71189559504783\\
0.0161559809843987	3.10399191153337\\
0.0316227766016838	3.23009526623321\\
0.061896581889126	3.23009634384508\\
0.121152765862859	3.23009620997335\\
0.237137370566166	3.23009593521808\\
0.464158883361278	3.23009566220585\\
0.908517575651686	3.23009413527419\\
1.77827941003892	3.23009434853905\\
3.48070058842841	3.2300948906304\\
6.81292069057961	3.23009511011919\\
13.3352143216332	3.23009563458988\\
26.1015721568254	3.23009468990958\\
51.0896977450692	3.23009587395819\\
100	3.23009598524218\\
};
\node[black,anchor=south west] at (axis cs:4e0,3.06){$\epsilon=0.01$};

\addplot [color=mycolor3,solid, forget plot]
  table[row sep=crcr]{%
1e-05	3.69273276052824\\
1.95734178148766e-05	3.69447811127284\\
3.83118684955728e-05	3.697261688282\\
7.49894209332456e-05	3.70262291474457\\
0.000146779926762207	3.71296904128744\\
0.000287298483335366	3.73388299818848\\
0.000562341325190349	3.77498063966422\\
0.00110069417125221	3.85706737769592\\
0.00215443469003188	4.02640166359567\\
0.00421696503428582	4.37632393595535\\
0.00825404185268018	5.14996977111069\\
0.0161559809843987	7.11901704015007\\
0.0316227766016838	7.98354039835244\\
0.061896581889126	7.98354401383944\\
0.121152765862859	7.98353107493953\\
0.237137370566166	7.98352622764507\\
0.464158883361278	7.98353105559035\\
0.908517575651686	7.98354083907565\\
1.77827941003892	7.98353074077898\\
3.48070058842841	7.98353437854478\\
6.81292069057961	7.98354600856873\\
13.3352143216332	7.98354758122917\\
26.1015721568254	7.98353732176234\\
51.0896977450692	7.98354280033106\\
100	7.98354349440077\\
};
\node[black,anchor=south west] at (axis cs:4e0,7.81){$\epsilon=0.02$};

\addplot [color=mycolor4,solid, forget plot]
  table[row sep=crcr]{%
1e-05	5.18598317017913\\
1.95734178148766e-05	5.18849285808392\\
3.83118684955728e-05	5.19556268717084\\
7.49894209332456e-05	5.22672980636617\\
0.000146779926762207	5.23099474673238\\
0.000287298483335366	5.27560654913596\\
0.000562341325190349	5.36626948110471\\
0.00110069417125221	5.54881354161445\\
0.00215443469003188	5.9438898066725\\
0.00421696503428582	6.83251331421466\\
0.00825404185268018	9.18574883881301\\
0.0161559809843987	20.2357661052605\\
0.0316227766016838	30.2137303597099\\
0.061896581889126	30.2138981815218\\
0.121152765862859	30.2138530502639\\
0.237137370566166	30.2140849592998\\
0.464158883361278	30.214074971216\\
0.908517575651686	30.2139497322001\\
1.77827941003892	30.2138760194343\\
3.48070058842841	30.2140245474673\\
6.81292069057961	30.214104651354\\
13.3352143216332	30.2145139782392\\
26.1015721568254	30.2139117840389\\
51.0896977450692	30.213976252577\\
100	30.2139830076914\\
};
\node[black,anchor=south west] at (axis cs:4e0,30){$\epsilon=0.025$};

\addplot [color=mycolor5,solid, forget plot]
  table[row sep=crcr]{%
1e-05	7.23823686831091\\
1.95734178148766e-05	7.24492807594648\\
3.83118684955728e-05	7.25760181978012\\
7.49894209332456e-05	7.28457290047892\\
0.000146779926762207	7.32349820251733\\
0.000287298483335366	7.41606465394899\\
0.000562341325190349	7.63589042062998\\
0.00110069417125221	8.0046371850535\\
0.00215443469003188	8.90969177317333\\
0.00421696503428582	11.2163213408786\\
0.00825404185268018	20.5771866318642\\
};
\node[black,anchor=west,yshift=-0.5mm] at (axis cs:0.00825,20.577){$\epsilon=0.0286$};

\addplot [color=mycolor5,mark=*,only marks, mark size=1,forget plot]
	table[row sep=crcr]{%
0.00825404185268018	20.5771866318642\\
};

\addplot [color=mycolor6,solid,forget plot]
  table[row sep=crcr]{%
1e-05	11.555949629017\\
1.95734178148766e-05	11.5819244993624\\
3.83118684955728e-05	11.6113974373344\\
7.49894209332456e-05	11.6762166714964\\
0.000146779926762207	11.8134547075316\\
0.000287298483335366	12.0919722181168\\
0.000562341325190349	12.6947969819322\\
0.00110069417125221	13.8386917498589\\
0.00215443469003188	16.923997642252\\
0.00421696503428582	28.9769454873407\\
};
\node[black,anchor=east,yshift=0.5mm] at (axis cs:0.004217,28.98){$\epsilon=0.0321$};

\addplot [color=mycolor6,mark=*,only marks, mark size=1,forget plot]
	table[row sep=crcr]{%
0.00421696503428582	28.9769454873407\\
};

\addplot [color=mycolor7,solid, forget plot]
  table[row sep=crcr]{%
1e-05	37.8458635566557\\
1.95734178148766e-05	37.9728080938108\\
3.83118684955728e-05	38.2232999254632\\
7.49894209332456e-05	38.7235234937772\\
0.000146779926762207	39.7424885540804\\
0.000287298483335366	41.9043577172274\\
0.000562341325190349	46.9090929591371\\
0.00110069417125221	61.3266905562967\\
0.00215443469003188	178.476190996559\\
};
\node[black,anchor=west,yshift=-0.5mm] at (axis cs:0.002154,178.48){$\epsilon=0.0357$};

\addplot [color=mycolor7,mark=*,only marks, mark size=1,forget plot]
	table[row sep=crcr]{%
0.00215443469003188	178.476190996559\\
};

\end{axis}
\end{tikzpicture}%
	\caption{Minimum of the $\mathscr{L}_2$-gain performance bound $\gamma$ with time-invariant $c$ ($\delta c_\mathrm{max}=0$) and time-varying $k$. The dots indicate the last point of feasibility.} \label{fig:exp4_DT}
\end{figure}
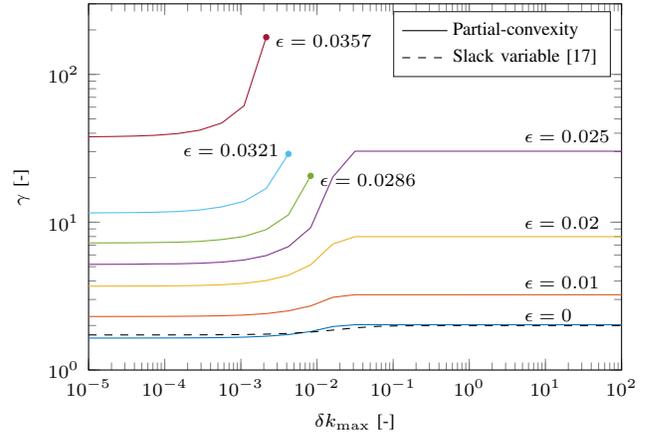

\begin{figure}
	\input{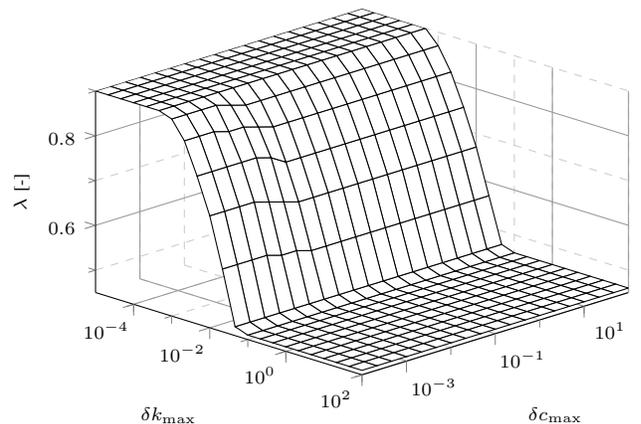}
	\caption{Maximum of $\lambda$ to guarantee AQS w.r.t. $\lambda\sP$ for a time-varying $k$ and $c$ with the partial-convexity method.} \label{fig:exp3_DT}
\end{figure}

In this section, the AQS and AQP results are demonstrated on the mass-spring-damper system from~\cite{Gahinet1996} with time-invariant and time-varying parameters. For the simulation example, Matlab 2014b with Yalmip and SeDuMi3 are used. The system is discretized using a first-order approximation~\cite{Toth2010a}:
\begin{equation}
\begin{aligned}
x(t+1) &= \Afnc_\mathrm{d}\left(f(t),c(t)\right) x(t) + \left[ \begin{array}{c} 0 \\ T_s \end{array} \right]u(t), \\
\Afnc_\mathrm{d}(k,c) &= I+T_s\Afnc_\mathrm{c}(k,c), \hspace{0.5cm} \Afnc_\mathrm{c}(k,c) \!=\!\! \left[\!\! \begin{array}{cc} 0 & \! 1 \\ -k & \! -c \end{array} \!\! \right]\!\!, \\
y(t) &= \left[ \begin{array}{cc} 1& 0 \end{array} \right] x(t),
\end{aligned}
\end{equation}
where $T_s=1/20$\,s is the sampling time and $k>0$, $c>0$ are the stiffness and the damping coefficients, which are assumed to be time-varying. The admissible trajectories of $k$ and $c$ are
\begin{equation} \label{eq:expKandC}
k(t) = k_0\bigl(1 + p_1(t) \bigr),\hspace{1cm} c(t) = c_0\bigl(1 + p_2(t) \bigr),
\end{equation}
where $k_0=1$, $c_0=1$, $\sP = [-1,1]\times[-1,1]$, and $\sV=[-\delta k_\mathrm{max},\delta k_\mathrm{max}]\times[-\delta c_\mathrm{max},\delta c_\mathrm{max}]$. The following three experiments are performed:
\begin{enumerate}
	\item For time-invariant $c$ ($\delta c_\mathrm{max}=0$) and time-varying $k$, the parameter box $\lambda\sP\coloneqq\{\lambda p:p\in\sP\}$ is uniformly expanded with $0<\lambda\leq1$ to find the maximum $\lambda$ for which stability is guaranteed. This is done with respect to values of $\delta k_\mathrm{max}\in\left[10^{-5},~10^2\right]$ and $\epsilon\in[0,1]$. The results are shown in Fig.~\ref{fig:exp2_DT}.
	\item For time-invariant $c$ ($\delta c_\mathrm{max}=0$) and time-varying $k$, the performance gain $\gamma$ is computed for $0.2\sP$. This is done w.r.t. $\delta k_\mathrm{max}\in\left[10^{-5},~10^2\right]$ and $\epsilon\in[0,1]$. The results are shown in Fig.~\ref{fig:exp4_DT}.
	\item For time-varying $k$ and $c$, the parameter box $\lambda\sP$ is expanded to find the maximum $\lambda$ for which stability is guaranteed. This is done w.r.t. $\delta k_\mathrm{max}\in\left[10^{-5},~10^2\right]$ and $\delta c_\mathrm{max}\in\left[10^{-4},~10^2\right]$. The results are shown in Fig.~\ref{fig:exp3_DT}.
\end{enumerate}
Note that the case of time-invariant $k$ ($\delta k_\mathrm{max}=0$) is not considered, as it is pointed out in~\cite{Gahinet1996}: `the stability region seems to be essentially determined by $\delta k_\mathrm{max}$'.

Fig.~\ref{fig:exp2_DT} shows that the slack variable method~\cite{Souza2006} is indeed more conservative (in line with \cite[Thm. 1]{Souza2006}). This difference is clearly visible for small $\delta k_\mathrm{max}$. For larger values of $\delta k_\mathrm{max}$, the maximum parameter box size coincides with the quadratic stability test ($\delta k_\mathrm{max}=\infty$), as expected. Furthermore, in the region $\delta k_\mathrm{max}\in[0.03,0.5]$, the slack variable method seems to outperform the proposed partial-convexity argument based method. Our method experiences numerical problems in the orange area for $\epsilon=0$, see Fig.~\ref{fig:exp2_DT}. In this area, neither feasibility or infeasibility can be concluded. Hence, for the proposed method, more research needs to be performed to improve numerical properties, but it is unclear if the slack variable method really outperforms the partial-convexity argument in that orange area. A similar result is also obtained for the $\mathscr{L}_2$-gain analysis in Fig.~\ref{fig:exp4_DT} (the area indicating numerical problems is not displayed in the figure).

Fig.~\ref{fig:exp2_DT} is similar to \cite[Fig. 1]{Gahinet1996} (CT case) in terms of the shape, as expected. However, the domain where the magnitude of $\lambda$ decreases is shifted by a factor $10$, which is due to the discretization of the model. Furthermore, Fig.~\ref{fig:exp3_DT} shows that the maximum size of the parameter box is almost independent on $\delta c_\mathrm{max}$, as has also been seen for the CT case \cite{Gahinet1996}.

\section{Conclusion} \label{sec:conclusion}
In this paper, we have proposed an LMI-based analysis for robust stability, robust state convergence, and robust performance of linear systems against uncertain and/or time-varying parameters. Affine quadratic stability is certified by finding a Lyapunov function, which is affine in the parameter. In order to get a tractable solution, the partial-convexity argument of~\cite{Gahinet1996} is extended for third-order terms to handle the discrete-time case. In our simulation example, the partial-convexity argument seems to be less conservative than the quadratic stability or the slack variable approach. For future research, the DT results can be extended to Lyapunov functions with quadratic dependence on the scheduling signal, similar to the CT case in~\cite{Trofino2001}. Alternatively, the DT result can be extended to the incremental stability framework for LPV systems with bounded rates of variations~\cite{Scorletti2015}.


\bibliographystyle{ieeetr}        
\bibliography{library.bib}           

\begin{thebibliography}{10}

\bibitem{Shamma1988}
J.~S. Shamma, {\em Analysis and design of gain scheduled control systems}.
\newblock PhD thesis, Massachusetts Institute of Technology, 1988.

\bibitem{Mohammadpour2012}
J.~Mohammadpour and C.~Scherer, eds., {\em Control of Linear Parameter Varying
  Systems with Applications}.
\newblock Springer, 2012.

\bibitem{Briat2015}
C.~Briat, {\em Linear Parameter-Varying and Time-Delay Systems: Analysis,
  Observation, Filtering \& Control}.
\newblock Springer, 2015.

\bibitem{Zhou1996}
K.~Zhou, J.~C. Doyle, and K.~Glove, {\em Robust and Optimal Control}.
\newblock Prentice hall, 1996.

\bibitem{Ebihara2015}
Y.~Ebihara, D.~Peaucelle, and D.~Arzelier, {\em S-Variable Approach to
  LMI-Based Robust Control}.
\newblock Springer London, 2015.

\bibitem{Megretski1997}
A.~Megretski and A.~Rantzer, ``System analysis via integral quadratic
  constraints,'' {\em IEEE Trans. on Automatic Control}, vol.~42, no.~6,
  pp.~819--830, 1997.

\bibitem{Helmersson1999}
A.~Helmersson, ``An {IQC}-based stability criterion for systems with slowly
  varying parameters,'' in {\em Proc. of the 14th IFAC World Congress},
  (Beijing, China), Jun. 1999.

\bibitem{Powers2001}
V.~Powers and B.~Reznick, ``A new bound for {P\'{o}lya's Theorem} with
  applications to polynomials positive on polyhedra,'' {\em J. of Pure and
  Applied Algebra}, vol.~164, pp.~221--229, 2001.

\bibitem{Molchanov2002}
A.~P. Molchanov and D.~Liu, ``Robust absolute stability of time-varying
  nonlinear discrete-time systems,'' {\em IEEE Trans. on Circuits and Systems
  I: Fundamental Theory and Applications}, vol.~49, no.~8, pp.~1129--1137,
  2002.

\bibitem{Koroglu2007}
H.~K\"{o}ro\u{g}lu and C.~W. Scherer, ``Robust performance analysis for
  structured linear time-varying perturbations with bounded
  rates-of-variation,'' {\em IEEE Trans. on Automatic Control}, vol.~52, no.~2,
  pp.~197--211, 2007.

\bibitem{Gahinet1996}
P.~Gahinet, P.~Apkarian, and M.~Chilali, ``{Affine parameter-dependent Lyapunov
  functions and real parametric uncertainty},'' {\em IEEE Trans. on Automatic
  Control}, vol.~41, no.~3, pp.~436--442, 1996.

\bibitem{Chilali1999}
M.~Chilali, P.~Gahinet, and P.~Apkarian, ``{Robust pole placement in LMI
  regions},'' {\em IEEE Trans. on Automatic Control}, vol.~44, no.~12,
  pp.~2257--2270, 1999.

\bibitem{Mahmoud2002}
M.~S. Mahmoud, ``Discrete-time systems with linear parameter-varying: stability
  vand $\mathcal{H}_\infty$,'' {\em J. of Mathematical Analysis and
  Applications}, vol.~269, no.~1, pp.~369--381, 2002.

\bibitem{Oliveira1999b}
M.~C. {de Oliveira}, J.~C. Geromel, and {Liu Hsu}, ``{LMI characterization of
  structural and robust stability: the discrete-time case},'' {\em Linear
  Algebra and its Applications}, vol.~296, pp.~27--38, 1999.

\bibitem{Blanchini1999}
F.~Blanchini and S.~Miani, ``A new class of universal {Lyapunov} functions for
  the control of uncertain linear systems,'' {\em IEEE Trans. on Automatic
  Control}, vol.~44, no.~3, pp.~641--647, 1999.

\bibitem{Daafouz2001}
J.~Daafouz and J.~Bernussou, ``Parameter dependent {Lyapunov} functions for
  discrete time systems with time varying parametric uncertainties,'' {\em
  Systems \& Control Letters}, vol.~43, no.~5, pp.~355--359, 2001.

\bibitem{Souza2006}
C.~E. {de Souza}, K.~A. Barbosa, and A.~T. Neto, ``Robust $\mathcal{H}_\infty$
  filtering for discrete-time linear systems with uncertain time-varying
  parameters,'' {\em IEEE Trans. on Signal Processing}, vol.~54, no.~6,
  pp.~2110--2118, 2006.

\bibitem{Amato2005}
F.~Amato, M.~Mattei, and A.~Pironti, ``Gain scheduled control for discrete-time
  systems depending on bounded rate parameters,'' {\em Int. J. of Robust and
  Nonlinear Control}, vol.~15, no.~11, pp.~473--494, 2005.

\bibitem{Iwasaki2001}
T.~Iwasaki and G.~Shibata, ``{LPV} system analysis via quadratic separator for
  uncertain implicit systems,'' {\em IEEE Trans. on Automatic Control},
  vol.~46, no.~8, pp.~1195--1208, 2001.

\bibitem{Peaucelle2007}
D.~Peaucelle, D.~Arzelier, D.~Henrion, and F.~Gouaisbaut, ``Quadratic
  separation for feedback connection of an uncertain matrix and an implicit
  linear transformation,'' {\em Automatica}, vol.~43, pp.~795--804, 2007.

\bibitem{Wei2008}
C.-P. Wei and L.~Lee, ``Stability analysis of discrete {LPV} systems subject to
  rate-bounded parameters,'' in {\em Proc. of the 17th IFAC World Congress},
  (Seoul, Korea), pp.~6383--6388, Jul. 2008.

\bibitem{Oliveira2001}
M.~C. {de Oliveira} and R.~E. Skelton, ``Stability tests for constrained linear
  systems,'' in {\em Perspectives in Robust Control} (S.~{Reza Moheimani},
  ed.), ch.~15, pp.~241--257, Springer London, 2001.

\bibitem{Scherer2006}
C.~W. Scherer, ``{LMI} relaxations in robust control,'' {\em European J. of
  Control}, vol.~12, pp.~3--29, 2006.

\bibitem{Veenman2016}
J.~Veenman, C.~W. Scherer, and H.~K\"{o}ro\u{g}lu, ``Robust stability and
  performance analysis based on integral quadratic constraints,'' {\em European
  J. of Control}, vol.~31, pp.~1--32, 2016.

\bibitem{Khalil2002}
H.~K. Khalil, {\em Nonlinear Systems}.
\newblock Prentice Hall, 3th~ed., 2002.

\bibitem{Rawlings2015}
J.~B. Rawlings and D.~Q. Mayne, {\em Model Predictive Control: Theory and
  Design}.
\newblock Nob Hill Publishing, fifth~ed., 2015.

\bibitem{Scherer2015}
C.~Scherer and S.~Weiland, ``Linear matrix inequalities in control.'' Lecture
  Notes for a course of the Dutch Institute of Systems and Control, Mar. 2015.

\bibitem{Toth2010a}
R.~T\'{o}th, {\em Modeling and Identification of Linear Parameter-Varying
  Systems}.
\newblock Springer, 2010.

\bibitem{Oliveira1999}
M.~C. {de Oliveira}, J.~Bernussou, and J.~C. Geromel, ``A new discrete-time
  robust stability condition,'' {\em Systems \& Control Letters}, vol.~37,
  no.~4, pp.~261--265, 1999.

\bibitem{Trofino2001}
A.~Trofino and C.~E. {de Souza}, ``Biquadratic stability for uncertain linear
  systems,'' {\em IEEE Trans. on Automatic Control}, vol.~46, no.~8,
  pp.~1303--1307, 2001.

\bibitem{Scorletti2015}
G.~Scorletti, V.~Fromion, and S.~{de Hillerin}, ``Toward nonlinear tracking and
  rejection using {LPV} control,'' in {\em Proc. of the 1st IFAC Workshop on
  Linear Parameter Varying Systems}, (Grenoble, France), pp.~13--18, Oct. 2015.

\end{thebibliography}

\appendix

\section{Proof Lemma~\ref{lem:multconv}} \label{app-sec-pf-lem:multconv}

For sufficiency, let $p^*=(p_1^*,\ldots,p_\NP^*)$ be the global maximizer of $\Lfnc(\centerdot)$ over $\sP$. Assume that $p^*_i$ is not at a vertex of the hyper-rectangle, i.e., $\underline p_i<p_i^*<\overline p_i$, then
\begin{multline} \label{eq:localPoly}
g(p_i) = \Lfnc(p^*_1,\ldots,p^*_{i-1},p_i,p^*_{i+1},\ldots,p^*_\NP) = \\
F_{0,i} + p_i F_{1,i} + p^2_i F_{2,i} + Q_{i,i,i} p^3_i,
\end{multline}
where
{\small%
\begin{align*}
F_{0,i} &\!=\!  Q_0 \!+\! \sum_{\substack{j=1 \\ j\neq i}}^{\NP} Q_j p^*_j \!+\! \sum_{\substack{j,k=1 \\ j,k\neq i}}^{\NP} Q_{j,k} p_j^*p_k^* \!+\!  \sum_{\substack{j,k,l=1 \\j,k,l\neq i}}^{\NP} Q_{j,k,l} p_j^*p_k^*p_l^*, \\
F_{1,i} &= \Big( Q_i+ \sum_{\substack{j=1\\ j\neq i}}^\NP (Q_{i,j}+Q_{j,i}) p^*_j+ \\[-3mm]
& \hspace{2cm} \sum_{\substack{j,k=1\\ j,k\neq i}}^\NP(Q_{i,j,k}+Q_{j,i,k}+Q_{j,k,i}) p^*_jp^*_k  \Big), \\
F_{2,i} &=  \Big( Q_{i,i}+ \sum_{\substack{j=1\\ j\neq i}}^\NP (Q_{j,i,i} + Q_{i,j,i} + Q_{i,i,j}) p^*_j \Big).
\end{align*}} \vskip -2mm \noindent
Since $p^*$ is the global maximizer of $\Lfnc(\centerdot)$, it holds that
\begin{equation} \label{apx-eq:maxg}
y^{\!\top}\!\! g(p_i^*) y \geq \max(y^{\!\top}\!\! g(\underline p_i) y, ~~ y^{\!\top}\!\! g(\overline p_i) y),
\end{equation}
where $y\neq0$.

On the other hand, Condition~\eqref{lem-eq:confNece} imposes convexity of $g(p_i)$ on $[\underline p_i,\overline p_i]$, i.e., $\frac{1}{2} \frac{\partial^2 g(p_i)}{\partial p_i^2}= F_{2,i}+3Q_{i,i,i}p_i\succeq0$ is implied by~\eqref{lem-eq:confNece}. Therefore its maximum on that interval is on the edges $\underline p_i,\overline p_i$. So,
\begin{equation} \label{apx-eq:convxg}
y^{\!\top}\!\! g(p_i^*) y \leq \max(y^{\!\top}\!\! g(\underline p_i) y, ~~ y^{\!\top}\!\! g(\overline p_i) y).
\end{equation}
Combining~\eqref{apx-eq:maxg} and~\eqref{apx-eq:convxg} leads to
\begin{equation}
y^{\!\top}\!\! g(p_i^*) y = \max(y^{\!\top}\!\! g(\underline p_i) y, ~~ y^{\!\top}\!\! g(\overline p_i) y).
\end{equation}
Concluding, the maximum of $g(\centerdot)$ is obtained at the edges $\{\underline p_i, \overline p_i\}$ of $p_i$ in case~\eqref{lem-eq:confNece} is satisfied. By repeating the same argument for each $i$ implies that the maximum of the function $\Lfnc(\centerdot)$ is at a vertex of the hyper-rectangle $\cnvxhullP$ when Condition~\eqref{lem-eq:confNece} is satisfied.

\vfill

\end{document}